\documentclass[letterpaper]{article} 
\usepackage{aaai2027}  
\usepackage[hyphens]{url}  
\usepackage{graphicx} 
\usepackage{natbib}  
\usepackage{caption} 
\usepackage{amsmath}
\usepackage{algorithm}
\usepackage{algorithmic}
\usepackage{booktabs}
\usepackage{tikz}
\usepackage{array}  
\usepackage{adjustbox}  
\usetikzlibrary{arrows.meta,positioning,fit,backgrounds,calc}

\title{Don't Regenerate, Debug: A Domain-Specific Agent for Repairing Near-Miss Accelerator Kernels}
\makeatletter
\author{%
    \global\aaai@corrmultitrue%
    Yansong~Sun\textsuperscript{\rm 1}\equalcontrib,
    Shenxiu~Wu\textsuperscript{\rm 1}\equalcontrib,
    Siyuan~Chen\textsuperscript{\rm 1},
    Runlin~Hou\textsuperscript{\rm 1},
    Junhao~Qiu\textsuperscript{\rm 1},
    Junming~Cao\textsuperscript{\rm 2},
    Shudi~Shao\textsuperscript{\rm 2},
    Zhichao~Lu\textsuperscript{\rm 1}\corresponding,
    Qingfu~Zhang\textsuperscript{\rm 1}\corresponding
}
\makeatother
\affiliations{
    \textsuperscript{\rm 1}Department of Computer Science, City University of Hong Kong\\
    \textsuperscript{\rm 2}Application Software Engineering Lab, Huawei Technologies Ltd.\\
    yansonsun3-c@my.cityu.edu.hk, zhichao.lu@cityu.edu.hk, qingfu.zhang@cityu.edu.hk
}

\begin{document}

\nocopyright
\maketitle

\begin{abstract}
Kernel generation for hardware accelerators such as GPUs and NPUs has become a proving ground for large language models (LLMs), and state-of-the-art systems raise correctness through pipelines that couple LLMs with agentic reinforcement learning and evolutionary search. Such pipelines generate, compile, and execute large numbers of candidate kernels, discarding most of them and forgoing the opportunity to distill failures into reusable knowledge. Many discarded candidates are \emph{near-miss} operators that compile and run but fail numerical validation; each embodies genuine domain knowledge and a nontrivial investment in LLM inference, cross-compilation, and hardware execution. We argue for a paradigm shift: rather than regenerate, \emph{debug}. Debugging is far more constrained than generating from scratch: the search space is small and feedback is dense. We present a domain-specific debug agent that addresses three core challenges in autonomous repair: mitigating knowledge scarcity through retrieved patterns and diagnostic instrumentation, ensuring integrity through anti-cheat detection and full-coverage evaluation, and controlling cost via convergence guards and bounded iteration. Debugging serves two complementary roles: it extends the capability frontier by recovering operators that repeated regeneration fails to produce, and it lowers cost per deliverable operator. Debug Pass@1 achieves 66.7\% versus Regenerate Avg Pass@1's 25.9\% and Regenerate Pass@3's 40.7\%, while consuming 92.8\% fewer tokens per success than three-trial regeneration. Component ablations show that the knowledge base drives recovery, while integrity gates reject 12.5--33.3\% of the successes the workflow itself accepted.
\end{abstract}

\section{Introduction}

Large language models have made significant strides in hardware accelerator kernel generation, yet a fundamental gap persists between compilation success and functional correctness. On CUDA, the best-represented GPU programming language in LLM pretraining corpora, state-of-the-art models compile above 90\% but are functionally correct only 49--86\% of the time, a gap that widens with task complexity \citep{cudabench2026,multikernelbench2025}. This gap is not a CUDA artifact, and it widens as training-data resources shrink: entire Triton task categories such as quantization achieve zero functional correctness despite non-trivial compilation rates \citep{wang2026kernelbenchx}; on Google's TPU/Pallas platform, models compile near-perfectly (up to 99.6\%) yet achieve only 2.8--8.4\% functional correctness \citep{multikernelbench2025}; and on lower-resource platforms such as Intel's SYCL and Moore Threads' MUSA, practical correctness requires either dozens of evolutionary iterations or specialized training \citep{wiedemann2026kernelfoundry,cheng2026musacoder}. Each compiled-but-failing kernel represents sunk investment in LLM inference, cross-compilation, and hardware execution, yet current pipelines insufficiently exploit these \emph{near-miss} candidates for systematic repair.

Each such \emph{near-miss} operator embodies correct API usage, reasonable tiling strategies, and valid memory management, representing genuine domain knowledge and nontrivial computational investment. We therefore argue for a complementary paradigm: rather than regenerate, \emph{debug}. Debugging a \emph{near-miss} operator is far more constrained than generation from scratch: structure fixed, failure localized, validation provides dense feedback at each iteration. Convergence therefore requires fewer trials and tokens, and each repair distills reusable patterns for future repairs, and the setting matches how production bugs are actually addressed, since they cannot be resolved by regenerating from scratch.

However, autonomous debug in low-resource domains faces three hazards. (1)~\textbf{Knowledge scarcity}: domains underrepresented in pretraining lack implicit debugging knowledge. Direct AscendC generation yields at most 2.5\% functional correctness \citep{multikernelbench2025}. (2)~\textbf{Integrity hazards}: agents under pressure can satisfy validation dishonestly through reward hacking, inadequate test coverage, or degeneration to slower fallbacks \citep{li2026cudabeaver,sarkar2026correctness,chatterjee2026proofwright,baronio2025kevin,lange2025towards}. (3)~\textbf{Cost control}: loops can cycle indefinitely without convergence \citep{zhou2026externalization}.

To address these hazards, an \textbf{engine-orchestrated architecture} is presented: the engine owns validation, forensics, and integrity checking; the agent operates only during repair with no acceptance authority. Five mechanisms address the hazards: knowledge base and diagnostic instrumentation (knowledge scarcity), anti-cheat detection and full-coverage evaluation (integrity), convergence guards (cost control). We ablate each in turn, and report where the evidence supports the design and where it does not.

We instantiate this on AscendC for Huawei's Ascend NPUs, where these hazards manifest acutely. Even DSL-guided approaches achieving 90.4\% on simple benchmarks \citep{ascendcraft2026} leave substantial residual failures: our 27 \emph{near-miss} operators from NPUKernelBench all passed CANNBot's sophisticated generation yet failed numerical validation. AscendC exemplifies low-resource domains (scarce data, strict hardware constraints such as explicit memory and queue sync, limited tooling) where the three hazards are empirical bottlenecks. AscendC thus serves as a representative hardest-case setting, and our findings suggest the approach may extend to other low-resource accelerators.

Experiments on this operator set show Debug Pass@1 reaching 66.7\%, meaning numerically correct without kernel bypass and backed by full-coverage evidence, versus Regenerate Avg Pass@1's 25.9\% and Regenerate Pass@3's 40.7\%, at 92.8\% lower token cost per success. Debug recovers 11 operators that all three regeneration trials fail to produce, and leads CANNBot's precision-debug mode by 22.2pp under a matched protocol. Our integrity gates, meanwhile, reject repairs the workflow itself had accepted: raw pass rate without them overstates deliverable success.

Our contributions: \begin{itemize}
\item We validate the debug-over-regenerate paradigm for low-resource kernels, demonstrating recovery of operators regeneration cannot reach while consuming fewer tokens.
\item We present an engine-orchestrated architecture with five mechanisms addressing knowledge scarcity, integrity hazards, and cost control.
\item Ablation separates recovery from integrity effects: removing the knowledge base lowers Pass@1 and raises tokens per success, while anti-cheat and full-coverage evaluation reject unsupported repairs. Trustworthy delivery, not raw pass rate, is the right objective.
\end{itemize}

\begin{table}[t]
\centering
\small
\begin{tabular}{@{}l@{}}
\toprule
\textbf{Baseline (fails validation)} \\
\midrule
\begin{minipage}[t]{0.92\columnwidth}
\begin{verbatim}
acc.SetValue(i,
  acc.GetValue(i)+acc.GetValue(i-1));
\end{verbatim}
\end{minipage} \\
\midrule
\textbf{Repaired (passes validation)} \\
\midrule
\begin{minipage}[t]{0.92\columnwidth}
\begin{verbatim}
float sum=acc.GetValue(i)
          +acc.GetValue(i-1);
half r=(half)sum;
acc.SetValue(i, (float)r);
\end{verbatim}
\end{minipage} \\
\bottomrule
\end{tabular}
\caption{Real AscendC cumsum operator from our precision tuning. The baseline's pure-FP32 accumulation diverges from the CANN reference; a three-line per-step FP16 rounding fix recovers correctness.}
\label{tab:example}
\end{table}

\section{Related Works}

\subsection{LLM-Based Code Generation and Debugging}
A growing body of work equips LLMs to debug rather than merely generate code. InspectCoder \citep{inspectcoder2025} uses dynamic analysis to drive program repair via interactive debugging, reporting substantial gains over static approaches. Agentless \citep{agentless2024} shows that a simple localize--repair--validate pipeline can rival more elaborate autonomous agents on SWE-bench. CT-Repair \citep{huang2026ctrepair} drives agentic repair with execution evidence, compressing runtime traces into queryable graphs for multi-agent localization on Defects4J. These works establish that structured feedback and validation are central to LLM debugging. We inherit both principles but move to the hardware-kernel domain, where the feedback is numerical deviation, the validation is on-device execution, and the evidence is a forensic trace of intermediate kernel state.

\begin{figure*}[!t]
\centering
\includegraphics[width=\textwidth]{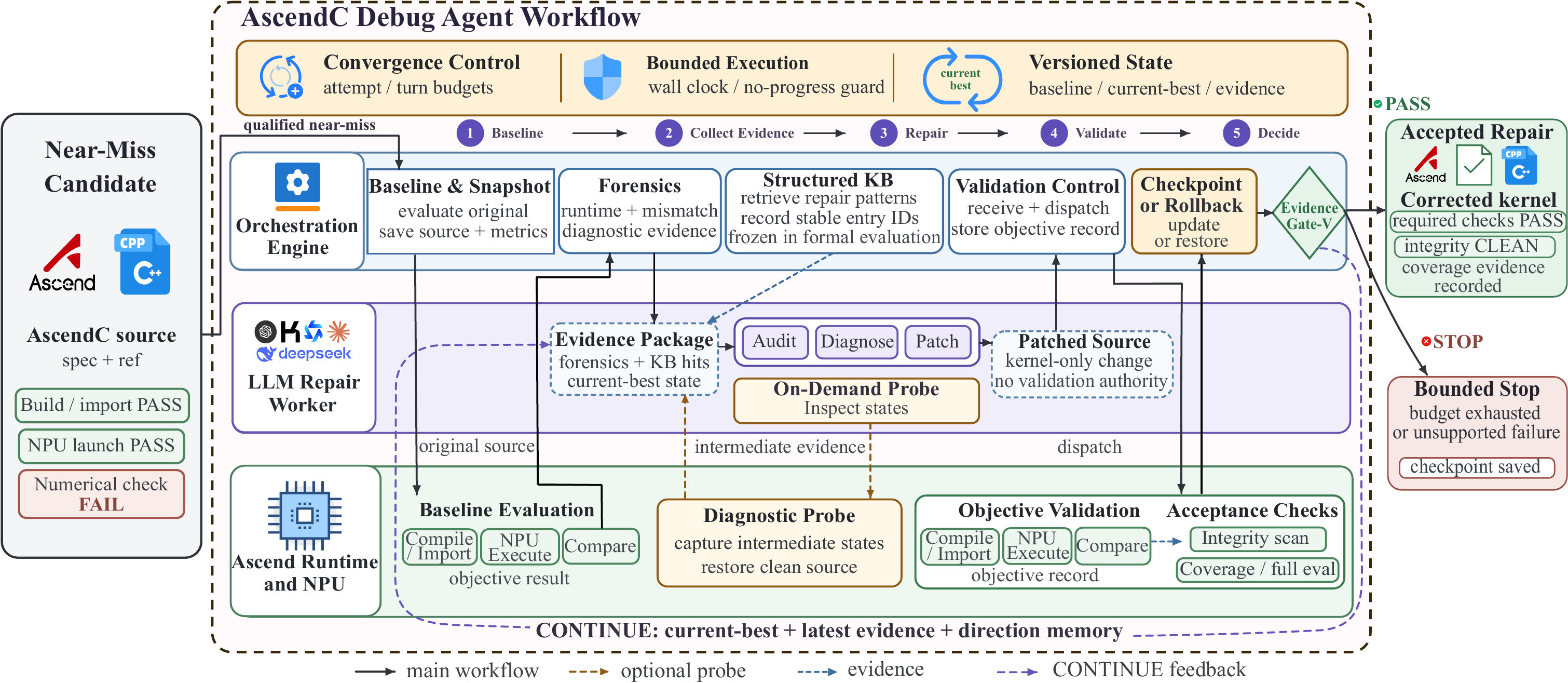}
\caption{AscendC Debug Agent overview. Given a \emph{near-miss} operator that compiles but fails numerical validation, the agent produces a repaired kernel through a three-stage loop (forensics $\rightarrow$ diagnose-and-fix $\rightarrow$ validate), with baseline entry and decision selection. The orchestration engine owns validation, forensics, and integrity checking, while five mechanisms (knowledge base, diagnostic instrumentation, anti-cheat detection, full-coverage evaluation, and convergence guards) guide and bound the loop.}
\label{fig:arch}
\end{figure*}

\subsection{Hardware Kernel Generation}
On the AscendC/Ascend platform, AscendCraft \citep{ascendcraft2026} achieves state-of-the-art results through DSL-guided transcompilation, in sharp contrast to the near-zero rate of direct generation. CANNBot \citep{cannbot2026}, Huawei's official agent suite for CANN development, bundles operator generation and precision-debug workflows; we adopt it as our baseline. Cross-platform benchmarks quantify the difficulty: MultiKernelBench \citep{multikernelbench2025} and NPUKernelBench \citep{npukernelbench2026} report Pass@1 below 2.5\% on AscendC, with API unfamiliarity as the dominant failure mode. On GPUs, GPU Kernel Scientist \citep{gpukernelscientist2025} and CudaForge \citep{cudaforge2025} explore iterative, feedback-driven optimization, and several systems already refine failing candidates in-loop by feeding compiler errors back between passes \citep{dai2026cuda,ascendcraft2026}, yet they stop at compilation and leave numerical failures unaddressed. CUDABeaver \citep{li2026cudabeaver} benchmarks automated CUDA debugging and documents ``repair by degeneration.'' SkillEvolver \citep{skillevolver2026} evolves a skill library with auditor checks against contamination from spurious successes, a concern our knowledge-base ingestion shares: we require full-coverage validation with high match rates, or explicit compatibility-mode acceptance when comprehensive evaluation is unavailable. We differ by repairing abandoned near-miss candidates post-hoc, certifying output with engine-owned validation and anti-cheat detection rather than scalar feedback alone.

These directions leave a gap at their intersection. Kernel-focused systems advance generation and optimization but treat numerically failing candidates as disposable; general-purpose repair fixes high-level code but faces none of the low-resource constraints, hardware-execution dependencies, or reward-hacking surface intrinsic to a validation-driven kernel loop. Our work occupies this intersection, repairing operators that compile and run yet fail numerical validation, and consequently must supply the anti-cheat and engine-owned validation that neither line of work requires.

\section{The AscendC Debug Challenge}

\subsection{The AscendC Programming Model}
AscendC is Huawei's C++ programming model for Ascend NPUs, exposing a tile-based, queue-driven execution model in which the developer explicitly manages data movement between Global Memory (GM) and the on-chip Unified Buffer (UB), tiling along problem dimensions, and a three-stage CopyIn--Compute--CopyOut pipeline synchronized through queues. Correctness depends on subtle interactions among dtype handling (e.g., FP16 accumulation precision), memory alignment, tiling boundaries, and API contracts that are poorly represented in LLM pretraining corpora. The combination of (1)~an unfamiliar low-level API, (2)~strict hardware constraints, and (3)~scarce training data explains both the near-zero one-shot success rate and, crucially, the prevalence of near-miss failures that motivate our approach.

\subsection{Problem Formulation}

We formalize the debug task as follows. Given a buggy AscendC kernel $K_{\text{buggy}} \leftarrow \text{LLM}(S)$ generated from specification $S$, which compiles successfully on hardware $h$ but fails numerical validation against reference implementation $R$ over test suite $\mathcal{T}$, a debug procedure maps it to a repaired kernel:
\begin{equation}
K_{\text{fixed}} = \operatorname{Debug}(K_{\text{buggy}}, S, R, \mathcal{T}, h;\; \mathcal{D}),
\end{equation}
where $K_{\text{buggy}}, S, R, \mathcal{T}, h$ define the problem instance, and $\mathcal{D}$ denotes the diagnostic evidence available to the agent (operator source, compilation logs, NPU execution traces, and numerical-deviation metrics). The output $K_{\text{fixed}}$ must satisfy, with $\tau$ the numerical tolerance:
\begin{itemize}
    \item $\text{Validate}(K_{\text{fixed}}, R, \mathcal{T}_{\text{ext}}, \tau) = \texttt{true}$ (correctness over the extended test suite);
    \item $K_{\text{fixed}} \models S$ (semantic consistency);
    \item $\text{Compile}(K_{\text{fixed}}, h) = \texttt{success}$ (hardware feasibility).
\end{itemize}

Table~\ref{tab:example} illustrates one representative instance. A generated cumsum kernel compiles, launches, and produces output of the correct shape; its results are comparable to the reference but do not clear the validation threshold. The root cause is subtle: the baseline accumulates partial sums in pure FP32, but the CANN reference uses FP16 arithmetic with per-step rounding. The numerical divergence compounds along the reduction axis. The fix introduces explicit per-step FP16 rounding to match the reference's arithmetic. Discarding this operator wastes all prior investment; recognizing and applying the precision-alignment pattern recovers it in a single repair.

\subsection{Why Autonomous Repair Is Hard}
Building such a system is non-trivial: on AscendC, the three hazards introduced above take domain-specific forms that a generic repair agent cannot handle. (1)~\textbf{Knowledge scarcity}: trained predominantly on CUDA corpora \citep{multikernelbench2025}, LLMs import idioms invalid on Ascend, such as assuming coalesced global-memory access where AscendC mandates explicit \texttt{DataCopy} over 32-byte-aligned UB tiles, or missing that FP16 accumulation must round per step to match the CANN reference. Such pitfalls are absent from pretraining corpora, so an unguided agent rediscovers them by trial and error. (2)~\textbf{Integrity hazards}: the Python-wrapper/C++-kernel split unique to the AscendC toolchain lets an agent route computation through the host wrapper (calling \texttt{at::} operators) so validation passes while the NPU kernel stays wrong, a bypass invisible to scalar pass/fail feedback. (3)~\textbf{Cost control}: precision tuning explores many near-identical dtype and rounding variants, so an unguarded loop readily oscillates or degrades a partially-correct kernel, burning scarce NPU cycles without progress.

\section{System Design}

As illustrated in Figure~\ref{fig:arch}, our system separates evidence generation from repair execution through a deterministic orchestration engine. The engine runs a three-stage loop of forensics (compiles and profiles the kernel), diagnosis-and-fix (invokes the LLM agent), and validation (re-executes to certify correctness), repeating until the kernel passes, a convergence guard triggers, or a budget limit is reached.

We deploy five mechanisms across the Guidance, Integrity, and Control layers, built on a foundational orchestration engine, to address the three hazards: knowledge retrieval and diagnostic instrumentation (Guidance Layer) mitigate knowledge scarcity; anti-cheat detection and full-coverage evaluation (Integrity Layer) counter integrity hazards; and convergence guards (Control Layer) enforce cost control. Each is ablated in turn, holding the remaining mechanisms fixed, so that each arm isolates one design claim for experimental validation.

\subsection{Orchestration Layer: Deterministic State Machine and Evidence Ownership}

The Orchestration Engine decides each transition deterministically from control state reconstructed out of a persistent event log, so every run is auditable and can resume from any interruption. Three phases belong to the engine rather than the agent: forensics generation, validation against the reference, and integrity checking on every candidate. The repair worker may run local diagnostics, but only engine-dispatched validation produces acceptance decisions.

\subsection{Guidance Layer: Knowledge Retrieval and Diagnostic Instrumentation}

Low-resource domains like AscendC lack the implicit debugging knowledge LLMs possess for well-represented languages. To compensate, we equip the agent with two guidance mechanisms that surface domain-specific patterns and localize failure root causes without requiring the LLM to rediscover them from scratch.

\paragraph{Structured knowledge base with auto-ingestion.} Debug patterns recur across operators: for instance, ``FP16 accumulation must round per step to match the CANN reference'' is a hardware-specific fact absent from pretraining corpora but applicable to any reduction kernel. We maintain a JSON-structured knowledge base retrieved deterministically after forensics, keyed by failure type and operator category (e.g., \texttt{precision\_failed} $\times$ \texttt{reduction}) so the agent sees only relevant patterns. Ingestion is quality-gated (Figure~\ref{fig:mechanisms}, top-left): only patterns with high match rates and clean integrity history are admitted, which does not certify complete correctness.

\paragraph{Diagnostic instrumentation.} Precision failures often present as diffuse numerical deviation that does not localize where divergence originates. The agent can then request that the engine instrument the kernel with multi-level tensor statistics at pipeline boundaries and re-run forensics. Because instrumentation costs agent turns and forensics cycles, the first iteration proceeds without it, and it applies only if that attempt fails (Figure~\ref{fig:mechanisms}, bottom-left).

\subsection{Integrity Layer: Anti-Cheat Detection and Full-Coverage Evaluation}

Agents under validation pressure can satisfy it dishonestly. We deploy two mechanisms that distinguish genuine repairs from reward-hacked ones.

\paragraph{Anti-cheat detection.} To catch the kernel-bypass hazard above, this integrity gate combines two checks, both recomputed by the engine on every validation: (1)~\emph{Wrapper validation} inspects the Python adapter for fallback computation patterns, detecting calls to reference operators or elimination of the custom kernel path; (2)~\emph{kernel-source scanning} detects forbidden host fallback (\texttt{at::}/\texttt{torch::} calls, tensor methods) and verifies the presence of a recognized custom-kernel launch site. Crucially, verdicts are recomputed against pre-iteration baseline snapshots. A repair is accepted only when this integrity verdict is clean (Figure~\ref{fig:mechanisms}, top-right). Bypassing repairs are marked \emph{wrapper-assisted} and rejected: numerically correct, but failing integrity checks.

\paragraph{Full-coverage evaluation.} A repair may pass a lightweight validation subset yet fail on the operator's complete shape/dtype/batch-size test matrix. To prevent false positives, we assign distinct roles to the two case sets NPUKernelBench provides per operator: the lightweight set drives fast iterative debugging, while the extended set (spanning the operator's full specification) certifies the final repair. When a repair passes the lightweight set, the engine escalates to full-coverage re-validation (Figure~\ref{fig:mechanisms}, bottom-right) and persists the resulting evidence (per-case pass/fail, numerical match rates, execution logs) as an immutable artifact. Only repairs backed by such full-coverage evidence, or whose lightweight set already \emph{is} the complete set, are certified to pass.

\subsection{Control Layer: Convergence Guards and Budget Enforcement}

Unbounded agent loops can oscillate between failed states, resubmit near-identical repairs, or degrade a partially working kernel, consuming scarce NPU cycles and LLM tokens without progress. The engine therefore owns every loop decision, and we separate what it enforces unconditionally from what it decides adaptively.

\paragraph{Hard bounds.} Four nested counters bound execution regardless of what the agent reports: 240 agent turns per session, 600 turns per task across all attempts, 5 repair rounds with per-failure-type sub-caps (3 for build errors, 5 for precision failures) so one error mode cannot exhaust the budget, and a wall-clock timeout. All are counted in the persistent event log, making them crash-safe, and all are expressed in turns rather than cost, so they do not drift when the backing model changes. Together with the validation gates, these bounds are a precondition for operating the harness at all: an agent able to negotiate its own termination could stop early on hard cases to inflate aggregate metrics. They are held fixed in every configuration we report, including all ablations, since removing them yields a non-terminating loop rather than a comparable arm.

\begin{figure*}[t]
\centering
\includegraphics[width=\textwidth]{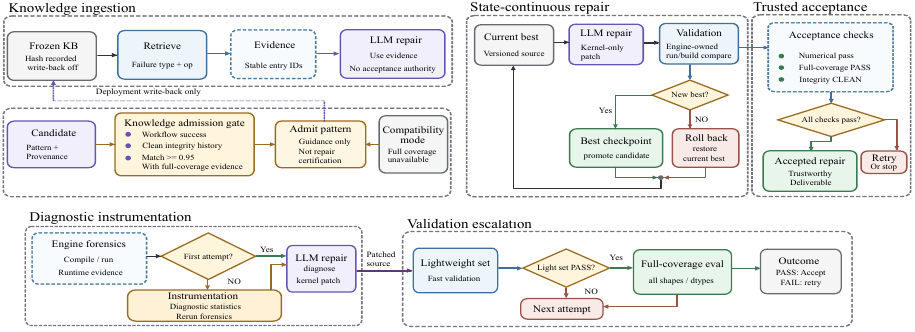}
\caption{Engine-owned mechanisms in detail. \textbf{Top-left}, knowledge base ingestion: a candidate pattern is admitted only when it reaches a 0.95 match rate under full-coverage evidence and a clean integrity history. \textbf{Top-middle}, state-continuous repair: the engine keeps the highest-scoring kernel and, after two consecutive non-improving attempts, rolls back to it. \textbf{Top-right}, trusted acceptance: a repair is accepted only if it passes the numerical, full-coverage, and integrity checks. \textbf{Bottom-left}, diagnostic instrumentation: the first attempt skips instrumentation, and only a failed attempt triggers tensor-statistic probes and re-profiling. \textbf{Bottom-right}, validation escalation: a lightweight pass triggers full-coverage evaluation across all shapes and dtypes.}
\label{fig:mechanisms}
\end{figure*}

\paragraph{State continuity.} The engine snapshots the highest-scoring kernel after each validation and atomically restores it when two consecutive attempts fail to beat it (Figure~\ref{fig:mechanisms}, top-middle), so repair resumes from the best-known state rather than the most recent one. It also injects prior fix types and verdicts into each prompt, forbidding directions already recorded as stagnant or regressed, and logs which retrieved knowledge-base entries were applied, yielding an auditable trail from evidence to fix.

\paragraph{Adaptive scheduling.} Within the hard bounds, a bundle of semantic rules decides when continuing is no longer productive: near-pass and fp16-ceiling detection, oscillation and same-direction stagnation detection, full-coverage non-improvement, and degenerate-round early stopping. A companion budget starts each task at 480 turns. The remaining hard allowance is released only when objective evidence shows real progress, namely a promoted best checkpoint or a full-coverage run passing 90\% of cases. A stalled task therefore never reaches the full allowance. This bundle is the one control component our ablation removes: unlike the hard bounds, it encodes judgments about progress and can be removed without breaking termination.

\section{Experiments}

\subsection{Experimental Setup}

\paragraph{Operator Cohort.} We evaluate on 27 frozen near-miss AscendC operators from NPUKernelBench \citep{npukernelbench2026}, a comprehensive benchmark for NPU kernel generation. These operators passed CANNBot's generation pipeline (compilation and execution) but failed numerical validation. The cohort spans L1 (8), L2 (9), L3 (6), and Advanced/L5--L7 (4) difficulty strata. All operators are held fixed across comparisons.

\paragraph{Evaluation Protocol.} A result passes only when it clears: (1)~complete test suite across all shapes/dtypes, (2)~numerical validation within tolerance, and (3)~integrity checks verifying no host-fallback bypass and the presence of a recognized custom-kernel launch site.

\paragraph{Methods Compared.} We compare Debug-Agent against CANNBot's \citep{cannbot2026} Regenerate mode. CANNBot is Huawei's official agent suite for CANN development, bundling state-of-the-art generation and precision-debug workflows with internal multi-round refinement. We run CANNBot in Regenerate mode: three independent trials per operator, each starting from fresh specification without access to prior trial outputs. Debug-Agent executes one repair task per operator, reusing the near-miss implementation from CANNBot's generation. We further compare against CANNBot's precision-debug mode, its public \texttt{ascendc-precision-debug} workflow. All experiments use Qwen3.8-Max-Preview (1M context), with an additional Kimi K3 run in the precision-debug comparison.

\subsection{Debug vs.\ Regenerate: Capability and Efficiency}

We investigate whether targeted repair recovers more operators than fresh regeneration (RQ1), and whether it reaches equivalent or better pass rates with fewer tokens (RQ2).

\paragraph{Metrics.} We report \textbf{Debug Pass@1} (single-run success), \textbf{Regenerate Avg Pass@1} (averaged over 81 runs: 27 operators $\times$ 3 trials), and \textbf{Regenerate Pass@3} (operator-level: any trial succeeds).

\subsubsection{RQ1: Does Debug Extend Recoverable Boundary?}

Table~\ref{tab:debug_vs_regen_strata} stratifies pass rates by task difficulty. Debug Pass@1 achieves 66.7\% (18/27) versus Regenerate Pass@3's 40.7\% (11/27), a +25.9pp advantage. The two sets overlap on only 7 operators: 11 succeed under Debug alone, and 4 under Regenerate alone. These results demonstrate that single-trial targeted repair extends the recoverable boundary beyond multi-round fresh regeneration from state-of-the-art pipelines.

\begin{table}[t]
\centering
\small
\setlength{\tabcolsep}{2.5pt}
\begin{tabular}{@{}lrrrrrr@{}}
\toprule
\textbf{Stratum} & \textbf{\(n\)} & \textbf{D\_Pass@1} & \textbf{R\_Avg@1} & \textbf{R\_Pass@3} & \textbf{$\Delta$pp} & \textbf{D/R} \\
\midrule
L1  & 8  & \textbf{75.0\%} & 12.5\% & 25.0\% & +50.0 & 5/1 \\
L2  & 9  & 33.3\%          & 11.1\% & 33.3\% & 0.0   & 2/2 \\
L3  & 6  & \textbf{83.3\%} & 38.9\% & 50.0\% & +33.3 & 3/1 \\
Adv & 4  & \textbf{100\%}  & 66.7\% & 75.0\% & +25.0 & 1/0 \\
\midrule
\textbf{Overall} & \textbf{27} & \textbf{66.7\%} & \textbf{25.9\%} & \textbf{40.7\%} & \textbf{+25.9} & \textbf{11/4} \\
\bottomrule
\end{tabular}
\caption{Capability by difficulty. D\_Pass@1 = Debug Pass@1, R\_Avg@1 = Regenerate Avg Pass@1, R\_Pass@3 = Regenerate Pass@3, $\Delta$pp = D\_Pass@1 minus R\_Pass@3 in percentage points, D/R = exclusive successes (Debug-only / Regen-only), Adv = Advanced (L5--L7).}
\label{tab:debug_vs_regen_strata}
\end{table}

\subsubsection{RQ2: Does Debug Achieve Higher Token Efficiency?}

Table~\ref{tab:debug_vs_regen_efficiency} reports token consumption normalized per operator and per success. Debug Pass@1 uses 0.247M uncached tokens/op versus Regenerate Avg Pass@1's 0.435M (1.76$\times$) and Pass@3's 1.305M (5.28$\times$). Per-success normalization amplifies these ratios: Debug requires 0.371M uncached tokens/success versus 1.678M (4.52$\times$) and 3.203M (8.63$\times$) for Regenerate, corresponding to 78\% and 88\% reductions. Including cache tokens, Debug achieves 86.3\% and 92.8\% cost reductions versus Regenerate Avg Pass@1 and Pass@3.

\begin{table}[t]
\centering
\small
\setlength{\tabcolsep}{2.5pt}
\begin{tabular}{@{}lccccc@{}}
\toprule
\textbf{Method} & \textbf{Pass@1} & \textbf{Unc/op} & \textbf{All/op} & \textbf{Unc/succ} & \textbf{All/succ} \\
\midrule
\textbf{D\_Pass@1} & \textbf{66.7\%} & \textbf{0.247M} & \textbf{18.96M} & \textbf{0.371M} & \textbf{28.44M} \\
R\_Avg@1  & 25.9\% & 0.435M & 53.98M & 1.678M & 208.22M \\
          &        & (1.76$\times$) & (2.85$\times$) & (4.52$\times$) & (7.32$\times$) \\
R\_Pass@3 & 40.7\% & 1.305M & 161.95M & 3.203M & 397.50M \\
          &        & (5.28$\times$) & (8.54$\times$) & (8.63$\times$) & (13.98$\times$) \\
\bottomrule
\end{tabular}
\caption{Efficiency by outcome. D\_Pass@1 = Debug Pass@1, R\_Avg@1 / R\_Pass@3 = Regenerate Avg/Pass@3, Unc = uncached tokens, All = all tokens (incl. cache), /op = per operator, /succ = per success. Parenthesized multipliers show cost relative to Debug. M = million tokens.}
\label{tab:debug_vs_regen_efficiency}
\end{table}

By reusing near-miss implementations through state-continuous repair, Debug-Agent recovers more operators than repeated fresh regeneration while significantly reducing token cost. Because Regenerate Avg Pass@1 is a run-level average rather than an operator-level binary outcome, a single McNemar test is undefined for it; we therefore compare Debug-Agent against each independent trial, obtaining exact two-sided $p=0.0129$, $0.0024$, and $0.0129$, all below the Bonferroni threshold of $0.05/3=0.0167$. These results demonstrate that targeted repair shifts the economic boundary of kernel failure recovery, turning kernel repair from a cost-prohibitive retry process into a viable production-scale strategy.

\subsection{Comparison with CANNBot Precision-Debug}

\textbf{RQ3: Does Debug Beat a Matched Precision-Debug Baseline?} We also compare Debug-Agent against CANNBot's precision-debug mode, the vendor's dedicated repair workflow, along the same two dimensions. Both methods start from identical frozen failing implementations and initial evidence, run under a matched budget on Qwen3.8-Max-Preview, and are judged by the same external full-eval and integrity protocol, yielding 27 paired Pass@1 outcomes.

Debug-Agent recovers more operators (Table~\ref{tab:cannbot}(a)): 18/27 (66.7\%) versus 12/27 (44.4\%), a +22.2pp overall margin, and the gap stays positive in all three sampled base strata (+37.5, +11.1, +33.3pp on L1--L3; both solve all four Advanced operators). Of the paired outcomes, 10 repairs are shared, 8 are Debug-only, and 2 are baseline-only. A supplementary Kimi K3 run yields a consistent +18.5pp margin (21/27 vs.\ 16/27); neither comparison reaches significance ($p=0.109$, $0.180$), but the direction holds across backbones.

On efficiency (Table~\ref{tab:cannbot}(b)), precision-debug uses 17.7\% fewer uncached tokens per operator (0.204M vs.\ 0.247M) but 1.98$\times$ more all-tokens, carrying a heavier cache load. Normalized by success, its uncached and all-token costs reach 1.24$\times$ and 2.97$\times$ those of Debug-Agent; Debug-Agent delivers each trustworthy repair with 66.3\% fewer all-tokens, achieving higher end-to-end token efficiency through a higher pass rate and lower cache burden.

\paragraph{Nominal Complexity vs.\ Empirical Difficulty.} Both agents solve all 4 Advanced operators, but achieve only 33.3\% and 22.2\% strict success on L2. Our coverage audit reveals that Advanced tasks contain only 5--10 fully exposed cases, whereas L2 tasks expand from 10 task-side cases to 49--58 full-evaluation cases with substantially broader shape coverage. Several L2 candidates fail on only one additional full-evaluation case. Although descriptive due to the small Advanced subset, this result shows that nominal structural complexity alone is insufficient to characterize debugging difficulty; evaluation coverage and task-to-full distribution shift must also be considered.

\begin{table}[t]
\centering
\small
\setlength{\tabcolsep}{3pt}
\begin{tabular*}{\columnwidth}{@{\extracolsep{\fill}}lrrrrr@{}}
\multicolumn{6}{@{}l}{\textbf{(a) Capability by difficulty}}\\
\toprule
\textbf{Stratum} & \textbf{\(n\)} & \textbf{D\_Pass@1} & \textbf{CBD\_Pass@1} & \textbf{$\Delta$pp} & \textbf{D/CBD}\\
\midrule
L1  & 8  & \textbf{75.0\%} & 37.5\%         & +37.5 & 3/0\\
L2  & 9  & \textbf{33.3\%} & 22.2\%         & +11.1 & 2/1\\
L3  & 6  & \textbf{83.3\%} & 50.0\%         & +33.3 & 3/1\\
Adv & 4  & \textbf{100\%}  & \textbf{100\%} & 0.0   & 0/0\\
\midrule
\textbf{Overall} & \textbf{27} & \textbf{66.7\%} & 44.4\% & \textbf{+22.2} & \textbf{8/2}\\
\bottomrule
\end{tabular*}
\vspace{2pt}
\begin{tabular*}{\columnwidth}{@{\extracolsep{\fill}}lccccc@{}}
\multicolumn{6}{@{}l}{\textbf{(b) Efficiency by outcome}}\\
\toprule
\textbf{Method} & \textbf{Pass@1} & \textbf{Unc/op} & \textbf{All/op} & \textbf{Unc/s} & \textbf{All/s}\\
\midrule
\textbf{D}   & \textbf{66.7\%} & 0.247M & \textbf{18.96M} & \textbf{0.371M} & \textbf{28.44M}\\
CBD & 44.4\% & \textbf{0.204M} & 37.53M & 0.458M & 84.45M\\
    &        & (0.82$\times$) & (1.98$\times$) & (1.24$\times$) & (2.97$\times$)\\
\bottomrule
\end{tabular*}
\caption{Debug-Agent (D) vs.\ CANNBot precision-debug (CBD) under a matched same-model repair protocol on 27 near-miss operators. \emph{(a)}~$\Delta$pp = D minus CBD; D/CBD = exclusive successes; Adv = L5--L7. \emph{(b)}~Unc/All = uncached/all tokens; /op, /s = per operator, per Pass@1 success; parenthesized values are CBD relative to D. M = million.}
\label{tab:cannbot}
\end{table}

\subsection{Ablation Study}

\textbf{RQ4: Which components make repair effective and trustworthy?} We ablate the five mechanisms on the same 27 operators, using Pass@1 as the success endpoint. Because the mechanisms target different failure modes, Table~\ref{tab:ablation} groups arms by each one's estimand rather than a single leaderboard: guidance is judged by recovery and token efficiency, integrity gates by how many workflow-accepted repairs they reject, and convergence control by long failures and token burden.

\begin{table}[t]
\centering
\small
\begin{tabular*}{\columnwidth}{@{\extracolsep{\fill}}lrrrrr@{}}
\multicolumn{6}{@{}l}{\textbf{(a) Guidance mechanisms}}\\
\toprule
\textbf{Config} & \textbf{Pass@1} & \textbf{$\Delta$pp} & \textbf{F/A} & \textbf{Unc/s} & \textbf{All/s}\\
\midrule
\textbf{Full} & \textbf{18/27} & -- & -- & 0.371M & \textbf{28.44M}\\
$-$KB & 13/27 & $-$18.5 & 8/3 & 0.490M & 45.01M\\
$-$Diagnostic & 15/27 & $-$11.1 & 4/1 & \textbf{0.357M} & 32.30M\\
\bottomrule
\end{tabular*}
\vspace{2pt}
\begin{tabular*}{\columnwidth}{@{\extracolsep{\fill}}lrrrr@{}}
\multicolumn{5}{@{}l}{\textbf{(b) Integrity gates}}\\
\toprule
\textbf{Gate removed} & \textbf{Arm-local} & \textbf{Pass@1} & \textbf{Unsup.} & \textbf{Risk}\\
\midrule
Anti-cheat & 16/27 & 14/27 & 2/16 & 12.5\%\\
Full-eval$^\dagger$ & 18/27 & 12/27 & 6/18 & 33.3\%\\
\bottomrule
\end{tabular*}
\vspace{2pt}
\begin{tabular*}{\columnwidth}{@{\extracolsep{\fill}}lrrrrr@{}}
\multicolumn{6}{@{}l}{\textbf{(c) Convergence control}}\\
\toprule
\textbf{Config} & \textbf{Pass@1} & \textbf{LF} & \textbf{Unc/op} & \textbf{All/op} & \textbf{All/s}\\
\midrule
Full & 18/27 & 4 & 0.247M & 18.96M & 28.44M\\
$-$Adapt.\ sched & 18/27 & 3 & \textbf{0.165M} & \textbf{13.60M} & \textbf{20.40M}\\
\bottomrule
\end{tabular*}
\caption{Component ablation on 27 AscendC near-miss operators, grouped by each arm's estimand. Pass@1 requires the full numerical suite plus a clean integrity review. \emph{F/A}: Full-only/Ablation-only Pass@1 successes. \emph{Arm-local}: successes the workflow accepts within the arm; \emph{Unsup.}: arm-local successes not upheld by external validation, i.e., false positives of the arm's own judgment; \emph{Risk}: Unsup./Arm-local. \emph{Unc}/\emph{All}: uncached/all tokens; \emph{/op}, \emph{/s}: per operator, per success. \emph{Adapt.\ sched}: adaptive scheduling. \emph{LF}: long failures, operators that consumed more turns or tokens than the Full arm's 75th percentile and still did not reach Pass@1, so they measure wasted effort rather than failure count. Hard turn/attempt bounds and validation gates are identical across all arms. Bold marks the best within each panel. $^\dagger$\texttt{no\_fulleval} is a matched shadow rebuilt from the Full trajectory at its first compact objective pass, not an independent run.}
\label{tab:ablation}
\end{table}

\paragraph{Guidance.} Removing the knowledge base drops Pass@1 from 18/27 to 13/27, with the paired direction favoring Full (8 operators succeed only with the KB, 3 only without). All tokens per success rise from 28.44M to 45.01M (1.58$\times$), the clearest evidence that retrieved domain knowledge improves both recovery and system-level token efficiency. Removing diagnostic evidence (engine forensics and instrumentation together) lowers Pass@1 to 15/27. Full spends more uncached tokens per operator, yet its higher success count yields lower all-tokens per success than the ablated arm (28.44M vs.\ 32.30M), suggesting diagnostic evidence trades single-task inference overhead for higher success yield through a mechanism that warrants further study.

\paragraph{Integrity.} Disabling anti-cheat leaves 16 arm-local successes, but 2 numerically passing repairs use wrapper-assisted computation, so only 14 reach Pass@1. Disabling full-coverage evaluation leaves 18 compact-set successes, of which 3 fail on the extended suite and 3 lack valid full-coverage evidence, leaving 12. The two gates thus reject 12.5\% and 33.3\% of arm-local successes: they are not merely audit logging but the mechanisms that keep locally-passing repairs from being reported as trustworthy.

\paragraph{Convergence control.} This arm removes the adaptive bundle only, that is, six semantic early-stopping rules together with the evidence-gated soft budget, so every task runs on the 600-turn hard ceiling from its first session. The per-session cap, attempt caps, wall-clock timeout, validation gates, and best-checkpoint rollback are identical in both configurations. Within those bounds, the bundle shows no measurable benefit on this cohort: Pass@1 is unchanged at 18/27, with no rise in long failures (4 to 3) or token cost. Since the hard bounds serve as a structural requirement rather than an empirically validated gain, we treat the present thresholds as requiring calibration. Because the arm removes the rules and the soft budget jointly, it does not attribute this result to either alone.

\section{Conclusion}
We argued that near-miss AscendC operators should be repaired rather than regenerated, and presented an engine-orchestrated debug agent that makes autonomous repair bounded and auditable. On 27 frozen near-miss operators, single-trial repair reached 66.7\% Pass@1 against 40.7\% for three regeneration trials at 92.8\% lower token cost per success, and outperformed each single regeneration trial under exact McNemar tests that survive Bonferroni correction.

Our ablation separates three roles. Retrieved domain knowledge is what drives recovery: removing the knowledge base costs five operators and raises all-tokens per success by 58\%, the clearest case of a mechanism that improves capability and cost together. The integrity gates instead govern what may be counted, rejecting 12.5\% and 33.3\% of the repairs the workflow itself had accepted; without them the rates above would have been reported higher, and wrong. The adaptive scheduling bundle we pre-registered showed no measurable benefit on this cohort, although the hard bounds it operates within remain a precondition for bounded execution.

The architecture rests on engine-owned validation, quality-gated knowledge accumulation, and deterministic bounds. These are properties of low-resource domains rather than of AscendC, though transfer remains untested. Our coverage audit further separates implementation complexity from evaluation-coverage complexity, a distinction we suggest future precision-debugging benchmarks report apart.

\bibliography{aaai2027}

\clearpage
\setlength{\belowcaptionskip}{7pt}
\setlength{\floatsep}{15pt plus 2pt minus 2pt}
\setlength{\dblfloatsep}{15pt plus 2pt minus 2pt}

\begin{center}
{\large\bfseries Appendix}\\[0.4em]
{\normalsize for ``Don't Regenerate, Debug: A Domain-Specific Agent\\for Repairing Near-Miss Hardware Operators''}
\end{center}
\vspace{0.8em}

This document provides the experimental protocols, per-operator outcomes, paired statistics, token-accounting details, final integrity results, and component-ablation evidence corresponding to RQ1--RQ4 in the main paper. All extended results use the same success definition, endpoints, and statistical units as the main paper.

\section{S1. Scope and Notation}

We use the following abbreviations throughout:

\begin{itemize}
\item \textbf{D}: Debug-Agent.
\item \textbf{R1--R3}: three independent fresh-regeneration trials.
\item \textbf{CBD}: CANNBot precision-debug.
\item \textbf{Q}: Qwen3.8-Max-Preview with a 1M-token context.
\item \textbf{K}: Kimi K3 with a 1M-token context.
\item \textbf{Unc. tokens}: input plus output tokens.
\item \textbf{All tokens}: input, output, cache-creation, and cache-read tokens.
\item \textbf{Pass@1}: one operator-level end-to-end task produces a candidate that passes the complete numerical evaluation and the final integrity criterion.
\end{itemize}

The statistical unit for method comparisons is an operator-level end-to-end task, not an internal repair attempt. Internal attempts share state and are not independent samples. For RQ3, the same 27 operators form repeated blocks in a balanced \(2\times2\) factorial design with method (D versus CBD) and model (Q versus K) as the two factors. Each operator contributes one end-to-end observation to each of the four cells; there is no additional technical replicate within an operator-cell.

\section{S2. Common Experimental Protocol}

\subsection{S2.1 Frozen cohort}

All primary experiments use the same 27 frozen near-miss AscendC operators. Each initial implementation had already compiled and executed in the upstream generation workflow but failed numerical validation. The cohort contains 8 L1, 9 L2, 6 L3, and 4 Advanced operators (two L5, one L6, and one L7). L5--L7 are combined only for display because their individual denominators are small.

The cohort and initial candidates were frozen before the compared outcomes were produced. No operator was removed based on a method's final result.

\subsection{S2.2 Pass@1 criterion}

Throughout this appendix, Pass@1 refers to one end-to-end task/run per operator. A task may contain multiple state-sharing diagnosis and repair rounds; it is not a single sampled completion. Pass@1 requires all of the following:

\begin{enumerate}
\item a materialized and executable candidate exists;
\item an external full-coverage evaluation completes;
\item every required numerical case passes the benchmark tolerance;
\item the final integrity check confirms a real path from \texttt{ModelNew.forward} through the extension binding, tiling, and custom AscendC kernel launch;
\item no framework/reference implementation performs the target computation, and no golden-output cache, monkey patch, dynamic Python execution, or case-manifest specialization substitutes for the kernel.
\end{enumerate}

Compact success or automatic anti-cheat \texttt{CLEAN} alone is insufficient. Attempt 0 omits probes unless NaN/Inf forensics triggers localization.

\subsection{S2.3 Final integrity criterion}

We distinguish the in-loop anti-cheat mechanism from the post-hoc integrity endpoint used for reported outcomes. One final criterion is applied to every full-eval passing candidate:

\begin{enumerate}
\item trace the executable path from the Python model to the custom kernel;
\item distinguish metadata/allocation calls from target computation;
\item reject host-side or framework-side implementations of the target semantics;
\item record a final integrity verdict for every candidate.
\end{enumerate}

Static checks first flag suspicious candidates. The current scanner assumes a fixed Python-wrapper, binding, and launch layout and does not fully cover direct kernel invocation, launches in \texttt{.asc} files, split registration and launch, or ATen calls used only for shape, dtype, allocation, tiling, and workspace management. Ambiguous flags are therefore resolved against source evidence under the fixed criterion above, applied identically to every method; automatic compatibility-aware anti-cheat verdicts are intermediate diagnostics rather than the reported endpoint. Under the automatic compatibility-aware anti-cheat verdict, 12 full-evaluation-passing candidates were initially classified as \texttt{CHEAT} but received a final \texttt{CLEAN} integrity verdict under the fixed source-evidence criterion (11 Regenerate runs and one Debug-Agent task), while one Debug-Agent candidate is confirmed as an integrity violation. The resulting Debug-Agent-minus-Regenerate Pass@3 gap is +25.9 pp, compared with +33.3 pp under the automatic compatibility-aware anti-cheat verdict.

\subsection{S2.4 Token accounting}

The primary cost endpoint is provider-reported token volume. For each operator, primary accounting covers the end-to-end task run that produces its reported terminal outcome; earlier runs terminated by infrastructure or provider interruptions are retained separately as stability records. Within the reported run, we aggregate input, output, cache-creation, and cache-read fields, deduplicate records by session, agent, and message identity, and include non-Pass@1 operators in the cost numerator. We report:

\begin{itemize}
\item tokens per attempted operator: total arm cost divided by 27;
\item tokens per Pass@1: total arm cost divided by the number of Pass@1 outcomes.
\end{itemize}

For Regenerate Pass@3, all three trials are included in the numerator. CLI pricing fields are subscription-credit proxies rather than auditable invoices and are therefore excluded from the paper's primary cost claims.

\subsection{S2.5 Statistical tests}

Paired binary outcomes use two-sided exact McNemar tests. Paired risk-difference intervals use 100,000 operator-level bootstrap resamples with a fixed seed. Marginal Wilson intervals are descriptive only. Because the cohort contains 27 operators, non-significant paired comparisons are reported as directional evidence.

\section{S3. RQ1--RQ2: Debug-Agent versus Fresh Regeneration}

\subsection{S3.1 Protocol}

Debug-Agent receives the frozen near-miss implementation and its failure evidence, and runs one state-continuous repair task per operator. Regenerate discards that candidate and runs three independent trials from the fresh operator specification. Regenerate trials use independent workspaces and do not observe other trial outputs. Both methods use Qwen3.8-Max-Preview and the same external full-eval and integrity endpoint.

This comparison deliberately evaluates two different recovery strategies: repairing sunk work versus discarding and regenerating it. Their internal control structures differ by design; the comparison is at the level of end-to-end effectiveness and cost under each declared budget.

\subsection{S3.2 Regeneration stability}

Tables~\ref{tab:supp-1} and~\ref{tab:supp-2} summarize regeneration stability at the operator and run levels, respectively.

\begin{table}[!t]
\centering
\footnotesize
\setlength{\tabcolsep}{4pt}
\begin{tabular}{@{}lrr@{}}
\toprule
Regenerate result & Operators & Fraction \\
\midrule
0/3 successful trials & 16 & 59.3\% \\
1/3 successful trials & 5 & 18.5\% \\
2/3 successful trials & 2 & 7.4\% \\
3/3 successful trials & 4 & 14.8\% \\
\bottomrule
\end{tabular}
\caption{Distribution of successful regeneration trials per operator.}
\label{tab:supp-1}
\end{table}

\begin{table}[!t]
\centering
\footnotesize
\setlength{\tabcolsep}{4pt}
\begin{tabular}{@{}lr@{}}
\toprule
Run-level endpoint & Pass@1 outcomes \\
\midrule
Regenerate trial 1 & 8/27 (29.6\%) \\
Regenerate trial 2 & 5/27 (18.5\%) \\
Regenerate trial 3 & 8/27 (29.6\%) \\
Regenerate Avg. Pass@1 & 21/81 (25.9\%) \\
Regenerate Pass@3 & 11/27 (40.7\%) \\
\bottomrule
\end{tabular}
\caption{Run-level regeneration endpoints. Avg. Pass@1 pools all 81 independent runs; Pass@3 counts an operator if any of its three trials passes.}
\label{tab:supp-2}
\end{table}

Only \texttt{BatchMatmul}, \texttt{MatmulTransB}, \texttt{scatter\_elements\_v2}, and \texttt{UnfoldGrad} succeed in all three regeneration trials. Repetition raises the number of operators solved from an average of 7 per trial to 11 under Pass@3, but does not eliminate trial variance.

\subsection{S3.3 Paired inference}

Regenerate Avg. Pass@1 is a run-level average over 81 runs and is not a 27-dimensional binary outcome. We therefore compare Debug-Agent separately with each regeneration trial; the resulting paired counts and exact tests are reported in Table~\ref{tab:supp-3}.

\begin{table*}[!t]
\centering
\footnotesize
\setlength{\tabcolsep}{4pt}
\begin{adjustbox}{max width=\textwidth}
\begin{tabular}{@{}lrrrrr@{}}
\toprule
Comparison & Both pass & D only & R only & Both fail & Exact \(p\) \\
\midrule
D vs. R1 & 6 & 12 & 2 & 7 & 0.0129 \\
D vs. R2 & 3 & 15 & 2 & 7 & 0.00235 \\
D vs. R3 & 6 & 12 & 2 & 7 & 0.0129 \\
\bottomrule
\end{tabular}
\end{adjustbox}
\caption{Paired exact McNemar comparisons between Debug-Agent and each regeneration trial.}
\label{tab:supp-3}
\end{table*}

All three values are below the Bonferroni threshold \(0.05/3=0.0167\). Each comparison pairs one Debug-Agent task with one regeneration trial. Because the same Debug-Agent outcome vector enters all three tests, they are three correlated views of one result rather than three independent replications.

The operator-level comparison against the complete three-trial Pass@3 budget has 7 shared successes, 11 Debug-only successes, 4 Regenerate-only successes, and 5 shared failures. The discordant 11:4 split gives a paired odds ratio of 2.75 and a two-sided exact McNemar \(p=0.1185\). The paired difference against each individual regeneration trial is significant after Bonferroni correction; the contrast against the complete three-trial Pass@3 budget is directionally consistent but not significant.

\subsection{S3.4 Complete per-operator matrix}

\texttt{P/x} denotes precision failure with \(x\%\) matched cases; \texttt{R/--} denotes runtime, import, or load failure without a valid match rate; \texttt{C/x} denotes a confirmed integrity violation. Pass denotes Pass@1.

Table~\ref{tab:supp-4} reports the complete operator-level outcomes underlying the aggregate comparisons.

\begin{table*}[!t]
\centering
\footnotesize
\setlength{\tabcolsep}{4pt}
\begin{tabular}{@{}p{0.05\textwidth}p{0.36\textwidth}cccc@{}}
\toprule
Level & Operator & D & R1 & R2 & R3 \\
\midrule
L1 & Add & Pass & P/78.0 & R/-- & R/-- \\
L1 & Cumsum & R/-- & P/43.1 & P/0.0 & R/-- \\
L1 & Sum & Pass & R/-- & R/-- & R/-- \\
L1 & LayerNorm & Pass & Pass & P/88.3 & Pass \\
L1 & AdamW & Pass & R/-- & R/-- & P/0.0 \\
L1 & EmbeddingDenseBackward & Pass & R/-- & R/-- & P/0.0 \\
L1 & NLLLoss & Pass & P/60.0 & R/-- & R/-- \\
L1 & IOU & P/53.3 & Pass & P/13.3 & P/13.3 \\
L2 & GroupNormSwish & Pass & P/94.0 & P/96.0 & Pass \\
L2 & MoeInitRouting & C/100.0 & P/0.0 & P/0.0 & R/-- \\
L2 & MoeFinalizeRouting & P/98.0 & R/-- & Pass & P/94.0 \\
L2 & DequantSwigluQuant & P/98.0 & R/-- & P/98.0 & Pass \\
L2 & FusedRopeWithQkNormAndKvCacheUpdate & P/0.0 & P/98.3 & P/98.3 & R/-- \\
L2 & HyenaFftSizePaddingRfft & Pass & R/-- & P/61.2 & P/30.0 \\
L2 & MaskedSoftmaxWithAttentionDropoutBackward & P/98.0 & P/52.9 & R/-- & P/0.0 \\
L2 & TanhGatedResidualAddBackward & Pass & P/74.0 & P/74.0 & P/82.0 \\
L2 & TimeDecayExponentialStabilization & P/52.0 & P/58.0 & P/74.0 & P/60.0 \\
L3 & BatchMatmul & P/64.7 & Pass & Pass & Pass \\
L3 & MatmulTransA & Pass & Pass & P/62.0 & R/-- \\
L3 & MatmulTransB & Pass & Pass & Pass & Pass \\
L3 & ConvStandard1d & Pass & R/-- & R/-- & R/-- \\
L3 & ConvStandard3d & Pass & P/0.0 & R/-- & P/0.0 \\
L3 & ConvDepthwise2d & Pass & P/0.0 & P/0.0 & P/0.0 \\
L5 & inplace\_index\_add\_with\_sorted & Pass & Pass & P/60.0 & Pass \\
L5 & scatter\_elements\_v2 & Pass & Pass & Pass & Pass \\
L6 & UnfoldGrad & Pass & Pass & Pass & Pass \\
L7 & NormRopeConcat & Pass & R/-- & P/50.0 & P/0.0 \\
\bottomrule
\end{tabular}
\caption{Complete per-operator outcome matrix for Debug-Agent and three regeneration trials. P/x is a precision failure with x percent matched cases; R/-- is a runtime, import, or load failure; C/x is a confirmed integrity violation.}
\label{tab:supp-4}
\end{table*}

The \texttt{Cumsum} example in Table 1 of the main paper comes from a separate precision-alignment demonstration batch, not from the frozen 27-operator cohort reported here. In this cohort, \texttt{Cumsum} is counted as a failure for D and all three Regenerate trials.

\subsection{S3.5 Failure and near-miss analysis}

Table~\ref{tab:supp-5} decomposes terminal outcomes into success, precision failure, runtime failure, and confirmed integrity violation.

\begin{table*}[!t]
\centering
\footnotesize
\setlength{\tabcolsep}{4pt}
\begin{adjustbox}{max width=\textwidth}
\begin{tabular}{@{}lrrrrr@{}}
\toprule
Method and unit & Pass@1 & Precision failure & Runtime failure & Confirmed integrity violation & Total \\
\midrule
Debug-Agent tasks & 18 & 7 & 1 & 1 & 27 \\
Regenerate runs & 21 & 36 & 24 & 0 & 81 \\
\bottomrule
\end{tabular}
\end{adjustbox}
\caption{Failure-category decomposition across Debug-Agent tasks and independent regeneration runs.}
\label{tab:supp-5}
\end{table*}

For an exploratory diagnosis, we define a near miss only among Pass@1 failures that execute enough full-eval cases to produce a match rate. Table~\ref{tab:supp-6} summarizes these valid failed outcomes.

\begin{table*}[!t]
\centering
\footnotesize
\setlength{\tabcolsep}{4pt}
\begin{adjustbox}{max width=\textwidth}
\begin{tabular}{@{}lrrrrrr@{}}
\toprule
Scope & Valid failed outcomes & Median match & IQR & At least 80\% & At least 90\% & At least 95\% \\
\midrule
D failures, excluding confirmed integrity violation & 7 & 64.7\% & 52.7--98.0\% & 3 & 3 & 3 \\
R Pass@3 failures, best failed trial & 14 & 51.5\% & 0.0--70.8\% & 2 & 1 & 1 \\
\bottomrule
\end{tabular}
\end{adjustbox}
\caption{Exploratory match-rate summary among failed outcomes with valid full-evaluation case results.}
\label{tab:supp-6}
\end{table*}

The regeneration row selects the best failure among up to three trials and therefore gives Regenerate an additional selection advantage. Two of the 16 Regenerate Pass@3 failures have only runtime failures and no meaningful match rate.

\subsection{S3.6 Full token decomposition}

Table~\ref{tab:supp-7} reports the complete token decomposition, and Table~\ref{tab:supp-8} presents the corresponding outcome-normalized efficiency measures.

\begin{table*}[!t]
\centering
\footnotesize
\setlength{\tabcolsep}{4pt}
\begin{tabular}{@{}lrrr@{}}
\toprule
Metric & D: 27 tasks & R: 81 trials & R/D \\
\midrule
Turns & 5,340 & 7,340 & 1.37x \\
Input tokens & 232,824 & 6,435,922 & 27.64x \\
Output tokens & 6,443,513 & 28,791,637 & 4.47x \\
Cache-creation tokens & 12,213,470 & 318,048,146 & 26.04x \\
Cache-read tokens & 493,011,671 & 4,019,255,358 & 8.15x \\
Uncached tokens & 6,676,337 & 35,227,559 & 5.28x \\
All tokens & 511,901,478 & 4,372,531,063 & 8.54x \\
\bottomrule
\end{tabular}
\caption{Full token decomposition. Uncached tokens are input plus output tokens; all tokens additionally include cache creation and cache reads.}
\label{tab:supp-7}
\end{table*}

\begin{table*}[!t]
\centering
\footnotesize
\setlength{\tabcolsep}{4pt}
\begin{adjustbox}{max width=\textwidth}
\begin{tabular}{@{}lrrrr@{}}
\toprule
Method and budget & Unc./operator & All/operator & Unc./Pass@1 & All/Pass@1 \\
\midrule
D Pass@1 & 0.247M & 18.96M & 0.371M & 28.44M \\
R Avg. Pass@1 & 0.435M & 53.98M & 1.678M & 208.22M \\
R Pass@3 & 1.305M & 161.95M & 3.203M & 397.50M \\
\bottomrule
\end{tabular}
\end{adjustbox}
\caption{Outcome-normalized token efficiency. Failed runs remain in the cost numerator.}
\label{tab:supp-8}
\end{table*}

The Pass@3 row includes all 81 trials, including failed trials. Compared with the complete three-trial regeneration budget, Debug-Agent uses 92.8\% fewer all tokens per Pass@1.

\section{S4. RQ3: Method \(\times\) Model Factorial Comparison with CANNBot Precision-Debug}

\subsection{S4.1 Balanced \(2\times2\) protocol}

The two factors are method \(\{\mathrm{D},\mathrm{CBD}\}\) and model \(\{\mathrm{Q},\mathrm{K}\}\), yielding four balanced cells: D-Q, D-K, CBD-Q, and CBD-K. All cells use the same frozen 27-operator cohort, starting implementations, attempt-0 evidence, repair-only constraint, 1M context, maximum effort, turn and task budgets, wall-time cap, full-evaluation endpoint, and the final integrity criterion. The execution environment and evaluation protocol are held fixed across cells; only the assigned method and model vary.

The complete factorial control protocol is summarized in Table~\ref{tab:supp-9}.

\begin{table*}[!t]
\centering
\footnotesize
\setlength{\tabcolsep}{4pt}
\begin{tabular}{@{}l@{\hspace{3.5em}}l@{}}
\toprule
Attribute & Factorial control \\
\midrule
Operator blocks & Same 27 frozen near-miss operators in all four cells \\
Methods & Debug-Agent; CANNBot precision-debug \\
Models & Qwen3.8-Max-Preview; Kimi K3 \\
Runs & One end-to-end run per operator-cell \\
Starting implementation & Identical frozen failing implementation \\
Initial evidence & Identical frozen attempt-0 evidence \\
Allowed behavior & Repair the existing project; fresh regeneration prohibited \\
Context / effort & 1M / max \\
Per-call cap & 240 turns \\
Task soft / hard cap & 480 / 600 turns \\
Repair attempts / operator task & At most 5 \\
Wall-time cap & 12 h per engine invocation (defensive ceiling) \\
Endpoint & Full-eval + final integrity criterion \\
\bottomrule
\end{tabular}
\caption{Controlled protocol for the balanced method-by-model comparison.}
\label{tab:supp-9}
\end{table*}

Infrastructure/provider interruptions retain the same workspace and session and are not reclassified as model failures.

The 12 h per-invocation timeout is a defensive ceiling against non-terminating sessions and did not determine any reported terminal outcome.

\subsection{S4.2 Qwen cell contrast}

Table~\ref{tab:supp-10} reports the paired capability comparison within the Qwen model cell.

\begin{table}[!t]
\centering
\footnotesize
\setlength{\tabcolsep}{4pt}
\begin{tabular}{@{}lr@{}}
\toprule
Statistic & Result \\
\midrule
Debug-Agent Pass@1 & 18/27 (66.7\%) \\
CANNBot precision-debug Pass@1 & 12/27 (44.4\%) \\
Observed paired risk difference & +22.2 pp \\
Paired bootstrap 95\% interval & [0.0, 44.4] pp \\
Both pass / both fail & 10 / 7 \\
D only / CBD only & 8 / 2 \\
Paired odds ratio & 4.00 \\
Exact McNemar \(p\) & 0.1094 \\
\bottomrule
\end{tabular}
\caption{Paired Qwen comparison between Debug-Agent and CANNBot precision-debug.}
\label{tab:supp-10}
\end{table}

The observed gap is +22.2 pp on this fixed cohort, with the paired bootstrap interval touching zero.

\subsection{S4.3 Factorial capability results}

We analyze all four cells symmetrically as an operator-blocked \(2\times2\) factorial experiment. The Qwen and Kimi cells differ only in the model factor, while D and CBD differ only in the method factor.

Table~\ref{tab:supp-11} reports Pass@1 by benchmark stratum, while Table~\ref{tab:supp-12} gives the pre-specified paired cell contrasts.

\begin{table}[!t]
\centering
\footnotesize
\setlength{\tabcolsep}{4pt}
\begin{adjustbox}{max width=\columnwidth}
\begin{tabular}{@{}lrrrrr@{}}
\toprule
Stratum & \(n\) & D-Q & D-K & CBD-Q & CBD-K \\
\midrule
L1 & 8 & 6/8 & 6/8 & 3/8 & 4/8 \\
L2 & 9 & 3/9 & 6/9 & 2/9 & 6/9 \\
L3 & 6 & 5/6 & 5/6 & 3/6 & 2/6 \\
Advanced & 4 & 4/4 & 4/4 & 4/4 & 4/4 \\
\textbf{Overall} & \textbf{27} & \textbf{18/27} & \textbf{21/27} & \textbf{12/27} & \textbf{16/27} \\
\bottomrule
\end{tabular}
\end{adjustbox}
\caption{Pass@1 across the balanced method-by-model cells, stratified by benchmark difficulty.}
\label{tab:supp-11}
\end{table}

\begin{table}[!t]
\centering
\footnotesize
\setlength{\tabcolsep}{4pt}
\begin{tabular}{@{}lrrr@{}}
\toprule
Comparison & Shared passes & A only / B only & Exact \(p\) \\
\midrule
D-Q vs. CBD-Q & 10 & 8 / 2 & 0.109 \\
D-K vs. CBD-K & 14 & 7 / 2 & 0.180 \\
D-K vs. D-Q & 17 & 4 / 1 & 0.375 \\
CBD-K vs. CBD-Q & 11 & 5 / 1 & 0.219 \\
\bottomrule
\end{tabular}
\caption{Pre-specified paired contrasts among the four method-by-model cells.}
\label{tab:supp-12}
\end{table}

For Pass@1, we define the method main effect as the average D-minus-CBD difference across the two models, the model main effect as the average K-minus-Q difference across the two methods, and the interaction as the difference between the two method contrasts:

\[
\Delta_{\mathrm{method}} =
\tfrac{1}{2}[(D_Q-CBD_Q)+(D_K-CBD_K)],
\]
\[
\Delta_{\mathrm{model}} =
\tfrac{1}{2}[(D_K-D_Q)+(CBD_K-CBD_Q)],
\]
\[
\Delta_{\mathrm{interaction}} =
(D_K-CBD_K)-(D_Q-CBD_Q).
\]

The resulting descriptive factorial effects are summarized in Table~\ref{tab:supp-13}.

\begin{table}[!t]
\centering
\footnotesize
\setlength{\tabcolsep}{4pt}
\begin{tabular}{@{}lr@{}}
\toprule
Factorial effect on Pass@1 & Estimate \\
\midrule
Method main effect: D minus CBD & \textbf{+20.4 pp} \\
Model main effect: K minus Q & \textbf{+13.0 pp} \\
Method \(\times\) model interaction & \textbf{-3.7 pp} \\
\bottomrule
\end{tabular}
\caption{Observed descriptive factorial effects on Pass@1.}
\label{tab:supp-13}
\end{table}

The positive method main effect reflects higher observed success for Debug-Agent after averaging over the two models. The positive model main effect reflects higher observed success with Kimi after averaging over the two methods. The observed interaction is small relative to both main effects: Debug-Agent leads by 22.2 pp under Qwen and 18.5 pp under Kimi, while Kimi improves D by 11.1 pp and CBD by 14.8 pp. Thus, the method advantage is directionally stable across models rather than being driven by one model cell.

The exact McNemar tests above quantify the four pre-specified paired cell contrasts; none alone reaches \(p<0.05\). The factorial estimates are descriptive operator-blocked effects on this fixed cohort. The model gain is concentrated on L2, and CBD decreases from 3/6 to 2/6 on L3.

\subsection{S4.4 Factorial token efficiency}

Table~\ref{tab:supp-14} reports token efficiency in all four cells, and Table~\ref{tab:supp-15} gives the within-model CANNBot-to-Debug-Agent ratios.

\begin{table*}[!t]
\centering
\footnotesize
\setlength{\tabcolsep}{4pt}
\begin{adjustbox}{max width=\textwidth}
\begin{tabular}{@{}lrrrrr@{}}
\toprule
Arm & Pass@1 & Unc./operator & All/operator & Unc./Pass@1 & All/Pass@1 \\
\midrule
D-Q & 18/27 & 0.247M & 18.96M & 0.371M & 28.44M \\
D-K & 21/27 & 0.376M & 7.505M & 0.483M & 9.649M \\
CBD-Q & 12/27 & 0.204M & 37.53M & 0.458M & 84.45M \\
CBD-K & 16/27 & 0.717M & 34.09M & 1.209M & 57.52M \\
\bottomrule
\end{tabular}
\end{adjustbox}
\caption{Token efficiency across the four method-by-model cells.}
\label{tab:supp-14}
\end{table*}

\begin{table*}[!t]
\centering
\footnotesize
\setlength{\tabcolsep}{4pt}
\begin{adjustbox}{max width=\textwidth}
\begin{tabular}{@{}lrrrr@{}}
\toprule
Model & CBD/D Unc./operator & CBD/D All/operator & CBD/D Unc./success & CBD/D All/success \\
\midrule
Qwen & 0.82x & 1.98x & 1.24x & 2.97x \\
Kimi & 1.91x & 4.54x & 2.50x & 5.96x \\
\bottomrule
\end{tabular}
\end{adjustbox}
\caption{Within-model CANNBot-to-Debug-Agent token ratios.}
\label{tab:supp-15}
\end{table*}

Within Qwen, CBD uses 17.7\% fewer uncached tokens per operator, while its all tokens per operator and all tokens per Pass@1 are 1.98x and 2.97x those of Debug-Agent, respectively. Under Kimi, the corresponding all-token ratios are 4.54x and 5.96x. The method-efficiency direction therefore repeats in both model strata. Provider cache accounting differs across models, so cross-model token totals are execution volumes rather than a common monetary measure; the model-stratified method ratios are the factorial efficiency contrasts used here.

\subsection{S4.5 Complete Qwen per-operator matrix}

\texttt{V/x} denotes an integrity violation, optionally with a numerical match rate. \texttt{P+V} and \texttt{R+V} denote a numerical/runtime failure with an additional integrity violation.

The complete Qwen operator-level comparison is reported in Table~\ref{tab:supp-16}.

\begin{table*}[!t]
\centering
\footnotesize
\setlength{\tabcolsep}{4pt}
\begin{tabular}{@{}p{0.05\textwidth}p{0.36\textwidth}ccp{0.19\textwidth}@{}}
\toprule
Level & Operator & D-Q & CBD-Q & Paired outcome \\
\midrule
L1 & Add & Pass & V/100.0 & D only \\
L1 & Cumsum & R/-- & R/-- & Both fail \\
L1 & Sum & Pass & Pass & Both pass \\
L1 & LayerNorm & Pass & P/13.3 & D only \\
L1 & AdamW & Pass & R/-- & D only \\
L1 & EmbeddingDenseBackward & Pass & Pass & Both pass \\
L1 & NLLLoss & Pass & Pass & Both pass \\
L1 & IOU & P/53.3 & P/13.3 & Both fail \\
L2 & GroupNormSwish & Pass & R/-- & D only \\
L2 & MoeInitRouting & V/100.0 & R/-- & Both fail \\
L2 & MoeFinalizeRouting & P/98.0 & P/98.0 & Both fail \\
L2 & DequantSwigluQuant & P/98.0 & P/80.0 & Both fail \\
L2 & FusedRopeWithQkNormAndKvCacheUpdate & P/0.0 & Pass & CBD only \\
L2 & HyenaFftSizePaddingRfft & Pass & Pass & Both pass \\
L2 & MaskedSoftmaxWithAttentionDropoutBackward & P/98.0 & P+V/45.1 & Both fail \\
L2 & TanhGatedResidualAddBackward & Pass & P/98.0 & D only \\
L2 & TimeDecayExponentialStabilization & P/52.0 & P/36.0 & Both fail \\
L3 & BatchMatmul & P/64.7 & Pass & CBD only \\
L3 & MatmulTransA & Pass & Pass & Both pass \\
L3 & MatmulTransB & Pass & Pass & Both pass \\
L3 & ConvStandard1d & Pass & P/0.0 & D only \\
L3 & ConvStandard3d & Pass & R+V/-- & D only \\
L3 & ConvDepthwise2d & Pass & P/0.0 & D only \\
L5 & inplace\_index\_add\_with\_sorted & Pass & Pass & Both pass \\
L5 & scatter\_elements\_v2 & Pass & Pass & Both pass \\
L6 & UnfoldGrad & Pass & Pass & Both pass \\
L7 & NormRopeConcat & Pass & Pass & Both pass \\
\bottomrule
\end{tabular}
\caption{Complete Qwen per-operator Pass@1 matrix for Debug-Agent and CANNBot precision-debug.}
\label{tab:supp-16}
\end{table*}

Among 13 numerically passing CBD-Q candidates, one (\texttt{Add}) materializes broadcasting in the framework adapter for 32/50 frozen cases and is excluded under the same conservative integrity rule used for Regenerate. The remaining 12 reach Pass@1. Three automatic anti-cheat \texttt{CHEAT} labels do not remain in the final results; all three candidates already failed full-eval, so this does not increase the success numerator. Every reference \texttt{model.py} remains byte-identical to the frozen input.

\subsection{S4.6 Complete Kimi Pass@1 matrix}

The corresponding Kimi operator-level comparison is reported in Table~\ref{tab:supp-17}.

\begin{table*}[!t]
\centering
\footnotesize
\setlength{\tabcolsep}{4pt}
\begin{tabular}{@{}p{0.05\textwidth}p{0.36\textwidth}ccp{0.19\textwidth}@{}}
\toprule
Level & Operator & D-K & CBD-K & Paired outcome \\
\midrule
L1 & Add & Pass & - & D only \\
L1 & Cumsum & - & - & Both fail \\
L1 & Sum & Pass & Pass & Both pass \\
L1 & LayerNorm & Pass & Pass & Both pass \\
L1 & AdamW & - & - & Both fail \\
L1 & EmbeddingDenseBackward & Pass & Pass & Both pass \\
L1 & NLLLoss & Pass & Pass & Both pass \\
L1 & IOU & Pass & - & D only \\
L2 & GroupNormSwish & Pass & Pass & Both pass \\
L2 & MoeInitRouting & - & - & Both fail \\
L2 & MoeFinalizeRouting & - & Pass & CBD only \\
L2 & DequantSwigluQuant & Pass & - & D only \\
L2 & FusedRopeWithQkNormAndKvCacheUpdate & Pass & Pass & Both pass \\
L2 & HyenaFftSizePaddingRfft & Pass & Pass & Both pass \\
L2 & MaskedSoftmaxWithAttentionDropoutBackward & Pass & - & D only \\
L2 & TanhGatedResidualAddBackward & Pass & Pass & Both pass \\
L2 & TimeDecayExponentialStabilization & - & Pass & CBD only \\
L3 & BatchMatmul & - & - & Both fail \\
L3 & MatmulTransA & Pass & Pass & Both pass \\
L3 & MatmulTransB & Pass & Pass & Both pass \\
L3 & ConvStandard1d & Pass & - & D only \\
L3 & ConvStandard3d & Pass & - & D only \\
L3 & ConvDepthwise2d & Pass & - & D only \\
L5 & inplace\_index\_add\_with\_sorted & Pass & Pass & Both pass \\
L5 & scatter\_elements\_v2 & Pass & Pass & Both pass \\
L6 & UnfoldGrad & Pass & Pass & Both pass \\
L7 & NormRopeConcat & Pass & Pass & Both pass \\
\bottomrule
\end{tabular}
\caption{Complete Kimi per-operator Pass@1 matrix for Debug-Agent and CANNBot precision-debug.}
\label{tab:supp-17}
\end{table*}

This matrix yields 14 shared passes, seven D-K-only passes, two CBD-K-only passes, and four shared failures, matching the paired counts in S4.3.

\section{S5. RQ4: Component Ablations}

\subsection{S5.1 Operational meaning of each arm}

Table~\ref{tab:supp-18} defines the mechanism removed and the controls retained in each ablation configuration.

\begin{table*}[!t]
\centering
\footnotesize
\setlength{\tabcolsep}{4pt}
\begin{tabular}{@{}p{0.14\textwidth}p{0.38\textwidth}p{0.38\textwidth}@{}}
\toprule
Configuration & Removed & Retained \\
\midrule
Full & Nothing & All five mechanisms and hard bounds \\
-KB & Structured retrieval/injection & Forensics, probes, gates, scheduling, hard bounds \\
-Diagnostic & Forensics, dependent KB search, and diagnostic tensor-statistic instrumentation & Validation, integrity gates, scheduling, hard bounds \\
-Adaptive scheduling & Six semantic early-stop rules and evidence-gated soft budget & 600-turn hard ceiling, per-call/attempt caps, timeout, validation, checkpoint/rollback \\
-Anti-cheat & Integrity acceptance gate & Numerical evaluation and all other mechanisms \\
-Full-eval\(^{\dagger}\) & Full-coverage acceptance gate at first compact pass & Compact objective result and integrity evidence available at that trajectory point \\
\bottomrule
\end{tabular}
\caption{Operational definition of each component-ablation configuration.}
\label{tab:supp-18}
\end{table*}

\texttt{-Diagnostic} removes the structured diagnostic-evidence capability as a unit: deterministic forensics, the KB search that consumes this evidence, and tensor-statistic instrumentation. Its estimand is the net effect of the whole capability; attribution to forensics, probing, or KB search individually is out of scope. \texttt{-Adaptive scheduling} is also a bundle. The hard execution bounds remain active, so this arm does not test unbounded execution.

\(^{\dagger}\) \texttt{-Full-eval} is a matched shadow reconstructed from the Full trajectory at its first compact objective pass; it is not an independent run.

\subsection{S5.2 Paired Pass@1}

Table~\ref{tab:supp-19} reports the paired Pass@1 comparison between Full and each ablation arm.

\begin{table*}[!t]
\centering
\footnotesize
\setlength{\tabcolsep}{4pt}
\begin{adjustbox}{max width=\textwidth}
\begin{tabular}{@{}lrrrrr@{}}
\toprule
Arm vs. Full & Pass@1 & Full only / Arm only & \(\Delta\) pp & Paired bootstrap 95\% interval & Exact \(p\) \\
\midrule
Full & 18/27 (66.7\%) & -- & -- & -- & -- \\
-KB & 13/27 (48.1\%) & 8 / 3 & -18.5 & [-40.7, +3.7] & 0.2266 \\
-Diagnostic & 15/27 (55.6\%) & 4 / 1 & -11.1 & [-25.9, +3.7] & 0.3750 \\
-Adaptive scheduling & 18/27 (66.7\%) & 4 / 4 & 0.0 & [-22.2, +22.2] & 1.0000 \\
-Anti-cheat & 14/27 (51.9\%) & 6 / 2 & -14.8 & [-33.3, +3.7] & 0.2891 \\
-Full-eval\(^{\dagger}\) & 12/27 (44.4\%) & 6 / 0 & -22.2 & [-37.0, -7.4] & 0.0312 \\
\bottomrule
\end{tabular}
\end{adjustbox}
\caption{Paired Pass@1 comparisons between Full and each ablation arm.}
\label{tab:supp-19}
\end{table*}

The first four real ablation comparisons all have intervals including zero and are read as directional on this cohort. The \texttt{-Full-eval} \(p\)-value is descriptive only because the shadow shares its trajectory with Full.

\subsection{S5.3 Token burden and long failures}

Table~\ref{tab:supp-20} reports outcome-normalized token burden and long-failure counts across configurations.

\begin{table*}[!t]
\centering
\footnotesize
\setlength{\tabcolsep}{4pt}
\begin{adjustbox}{max width=\textwidth}
\begin{tabular}{@{}lrrrrr@{}}
\toprule
Configuration & Unc./operator & All/operator & Unc./Pass@1 & All/Pass@1 & Long failures \\
\midrule
Full & 0.247M & 18.96M & 0.371M & 28.44M & 4 \\
-KB & 0.236M & 21.67M & 0.490M & 45.01M & 5 \\
-Diagnostic & 0.198M & 17.94M & 0.357M & 32.30M & 4 \\
-Adaptive scheduling & 0.165M & 13.60M & 0.247M & 20.40M & 3 \\
-Anti-cheat & 0.169M & 14.85M & 0.325M & 28.64M & 3 \\
-Full-eval\(^{\dagger}\) & 0.213M & 16.41M & 0.480M & 36.92M & 5 \\
\bottomrule
\end{tabular}
\end{adjustbox}
\caption{Token burden and long-failure counts across component-ablation configurations.}
\label{tab:supp-20}
\end{table*}

A long failure is a non-Pass@1 outcome whose burden reaches at least two of the three Full-arm 75th-percentile thresholds for turns, all tokens, and the recorded cost proxy. The Full-arm thresholds are frozen and applied unchanged to every arm.

Removing the KB lowers raw uncached tokens per attempted operator slightly but reduces Pass@1 enough that all tokens per Pass@1 rise by 58\%. This is the clearest observed capability-efficiency alignment. By contrast, the current adaptive-scheduling bundle shows no measured benefit: Pass@1 is unchanged, long failures do not increase, and observed token burden is lower without it. Because the arm jointly removes semantic rules and the soft budget, and execution windows are not perfectly matched, this result indicates that the current thresholds require calibration.

\subsection{S5.4 Common-success cost sensitivity}

To reduce confounding from different success sets, we compare only operators that reach Pass@1 under both Full and the corresponding arm. Table~\ref{tab:supp-21} reports this common-success sensitivity analysis.

\begin{table*}[!t]
\centering
\footnotesize
\setlength{\tabcolsep}{4pt}
\begin{adjustbox}{max width=\textwidth}
\begin{tabular}{@{}lrrrr@{}}
\toprule
Arm vs. Full & Common Pass@1 & Mean turns & Mean all tokens & Mean cost proxy \\
\midrule
-KB & 10 & +23.0\% & +27.4\% & +15.5\% \\
-Diagnostic & 14 & -14.4\% & +9.1\% & -5.3\% \\
-Adaptive scheduling & 14 & -28.0\% & -25.6\% & -29.6\% \\
-Anti-cheat & 12 & -30.6\% & -37.8\% & -39.4\% \\
-Full-eval\(^{\dagger}\) & 12 & 0.0\% & 0.0\% & 0.0\% \\
\bottomrule
\end{tabular}
\end{adjustbox}
\caption{Common-success cost sensitivity. Each row is restricted to operators that pass in both Full and the compared arm.}
\label{tab:supp-21}
\end{table*}

Positive values mean that the ablation arm consumes more. Successful subsets and execution windows differ across arms, and the integrity arms change acceptance rather than the repair computation alone, so these figures are sensitivity statistics.

\subsection{S5.5 Complete Pass@1 matrix}

\texttt{P} denotes Pass@1; \texttt{-} denotes failure or insufficient evidence.

Table~\ref{tab:supp-22} reports the complete operator-level Pass@1 matrix across Full and the five ablations.

\begin{table*}[!t]
\centering
\footnotesize
\setlength{\tabcolsep}{4pt}
\begin{tabular}{@{}p{0.35\textwidth}cccccc@{}}
\toprule
Operator & Full & -KB & -Diag. & -Sched. & -AC & -FE\(^{\dagger}\) \\
\midrule
L1-003 Add & P & P & P & P & P & P \\
L1-005 Cumsum & - & - & - & - & - & - \\
L1-007 Sum & P & P & P & P & P & P \\
L1-010 LayerNorm & P & P & P & P & P & P \\
L1-017 AdamW & P & P & P & P & - & - \\
L1-024 EmbeddingDenseBackward & P & P & P & P & P & P \\
L1-025 NLLLoss & P & - & - & P & - & - \\
L1-031 IOU & - & - & P & P & P & - \\
L2-002 GroupNormSwish & P & - & - & - & P & - \\
L2-005 MoeInitRouting & - & - & - & - & - & - \\
L2-006 MoeFinalizeRouting & - & P & - & P & - & - \\
L2-011 DequantSwigluQuant & - & - & - & - & - & - \\
L2-020 FusedRopeWithQkNormAndKvCacheUpdate & - & P & - & P & P & - \\
L2-023 HyenaFftSizePaddingRfft & P & P & P & - & P & P \\
L2-025 MaskedSoftmaxWithAttentionDropoutBackward & - & P & - & P & - & - \\
L2-029 TanhGatedResidualAddBackward & P & - & - & P & - & - \\
L2-030 TimeDecayExponentialStabilization & - & - & - & - & - & - \\
L3-001 BatchMatmul & - & - & - & - & - & - \\
L3-004 MatmulTransA & P & P & P & P & P & P \\
L3-005 MatmulTransB & P & - & P & P & P & - \\
L3-006 ConvStandard1d & P & - & P & - & P & P \\
L3-008 ConvStandard3d & P & - & - & P & - & - \\
L3-009 ConvDepthwise2d & P & - & P & P & P & P \\
L5-005 inplace\_index\_add\_with\_sorted & P & P & P & P & P & P \\
L5-010 scatter\_elements\_v2 & P & - & P & - & - & P \\
L6-021 UnfoldGrad & P & P & P & P & P & P \\
L7-004 NormRopeConcat & P & P & P & P & - & P \\
\bottomrule
\end{tabular}
\caption{Complete per-operator Pass@1 matrix for Full and five component ablations.}
\label{tab:supp-22}
\end{table*}

Non-monotonic per-operator outcomes are expected under stochastic repair. Aggregate differences must therefore be interpreted together with discordant pairs and uncertainty intervals.

\subsection{S5.6 Integrity-gate decomposition}

Table~\ref{tab:supp-23} separates arm-local acceptance from the final Pass@1 endpoint for the two integrity-gate ablations.

\begin{table*}[!t]
\centering
\footnotesize
\setlength{\tabcolsep}{4pt}
\begin{adjustbox}{max width=\textwidth}
\begin{tabular}{@{}lrrrrr@{}}
\toprule
Gate removed & Arm-local successes & Pass@1 & Confirmed rejection & Missing valid full-coverage evidence & Unsupported-success risk \\
\midrule
Anti-cheat & 16 & 14 & 2 wrapper-assisted candidates & 0 & 2/16 (12.5\%) \\
Full-eval\(^{\dagger}\) & 18 & 12 & 3 extended-suite failures & 3 full-eval crashes at the shadow point & 6/18 (33.3\%) \\
\bottomrule
\end{tabular}
\end{adjustbox}
\caption{Integrity-gate decomposition. Arm-local success is the success count under the ablated arm's own acceptance rule.}
\label{tab:supp-23}
\end{table*}

The two anti-cheat rejections are \texttt{ConvStandard3d}, which performs ATen CPU preprocessing/padding, and \texttt{scatter\_elements\_v2}, which uses ATen for data-dependent shape inference.

At the first compact objective pass in the Full trajectory, \texttt{MoeInitRouting}, \texttt{MatmulTransB}, and \texttt{ConvStandard3d} fail the extended suite. \texttt{NLLLoss}, \texttt{GroupNormSwish}, and \texttt{DequantSwigluQuant} lack a valid full-coverage result at that shadow point because the corresponding full-eval crashes. All three later obtain valid post-hoc full-coverage evidence in the primary Full experiment. They are counted as unsupported at the shadow point rather than as confirmed numerical failures.

\section{S6. Evaluation Coverage and Integrity Audit}

\subsection{S6.1 Evaluation coverage versus nominal level}

The four Advanced operators expose only 5--10 full-evaluation cases, while several L2 operators expand from 10 task-side cases to 49--58 full-evaluation cases with broader shape coverage. Several L2 candidates fail on only one additional full-eval case. Therefore, benchmark level and implementation complexity do not fully capture evaluation difficulty; task-to-full coverage shift is an additional axis.

\subsection{S6.2 Primary-result completeness}

Table~\ref{tab:supp-24} summarizes evidence completeness for the primary Debug-Agent and Regenerate results.

\begin{table}[!htbp]
\centering
\footnotesize
\setlength{\tabcolsep}{4pt}
\begin{tabular}{@{}>{\raggedright\arraybackslash}p{0.46\columnwidth}>{\centering\arraybackslash}p{0.23\columnwidth}>{\centering\arraybackslash}p{0.23\columnwidth}@{}}
\toprule
Check & Debug-Agent Qwen & Regenerate Qwen \\
\midrule
Frozen operators & 27/27 & 27/27 \\
Terminal tasks/runs & 27/27 & 81/81 \\
Materialized candidates & 27/27 & 81/81 \\
Independent post-hoc full-eval & 27/27 & 81/81 \\
Final integrity verdict & 27/27 & 81/81 \\
Cross-arm contamination detected & 0 & 0 \\
\bottomrule
\end{tabular}
\caption{Primary-result evidence completeness for Debug-Agent and regeneration.}
\label{tab:supp-24}
\end{table}

For Debug-Agent Qwen, 19 candidates pass full-eval, 18 reach final Pass@1, and one is excluded. For Regenerate, 21 runs pass full-eval and all 21 reach final Pass@1. The same final integrity criterion is applied to both methods.

\section{S7. Reproducibility Checklist}

The reproducibility record for these results is organized around the following artifacts:

\begin{enumerate}
\setlength{\itemsep}{0pt}
\setlength{\parskip}{0pt}
\setlength{\parsep}{0pt}
\item a frozen 27-operator manifest with difficulty level, compact/full case counts, coverage-equivalence status, and compact/full case-set hashes;
\item redacted model, context, effort, turn, attempt, timeout, and evaluator configuration;
\item a run manifest mapping each candidate ID to its result row and task/trial identity;
\item post-hoc full-eval summaries, automatic compatibility-aware anti-cheat verdicts, final integrity verdicts for every candidate, and available source-level resolution metadata;
\item per-candidate token aggregates and aggregation/deduplication code;
\item paired-outcome, bootstrap, Wilson-interval, and exact-McNemar scripts;
\item the derived no-full-eval shadow script, including the first compact-pass selection rule;
\item a script reproducing the reported headline counts and statistics from these structured inputs.
\end{enumerate}

All 27 Full tasks have terminal status, post-hoc numerical evidence, token aggregates, and final integrity verdicts. 

The packaged artifacts satisfy the following consistency checks:

\begingroup
\scriptsize
\begin{itemize}
\setlength{\itemsep}{0pt}
\setlength{\parskip}{0pt}
\item Debug-Agent Qwen: 18 Pass@1 outcomes
\item Regenerate Qwen: 21/81 run successes; 11/27 Pass@3
\item CANNBot precision-debug Qwen: 12 Pass@1 outcomes
\item Full / -KB / -Diagnostic / -Adaptive scheduling: 18 / 13 / 15 / 18
\item No duplicate task-trial identifiers
\item All primary candidates have full-eval and integrity evidence
\item No provider credentials or absolute infrastructure paths
\end{itemize}
\endgroup

\section{S8. Limitations and Claim Boundaries}

\begin{enumerate}
\item The cohort contains 27 near-miss operators from one NPU programming ecosystem. Transfer to other hardware and languages remains untested.
\item Debug-Agent has one end-to-end task per operator, whereas Regenerate has three independent trials. Internal Debug attempts share state and cannot be reinterpreted as independent samples.
\item Debug-Agent versus Regenerate Pass@3 and Debug-Agent versus CBD do not reach \(p<0.05\) under operator-level exact McNemar tests. Their paired advantages are directional observations on this cohort.
\item L5--L7 contain only four operators. Their 100\% observed success is descriptive and does not establish a general advantage on advanced tasks.
\item RQ3 is a balanced, operator-blocked method \(\times\) model factorial experiment with a fixed environment and one run per operator-cell. The 27 operators provide the matched blocks, but there is no additional technical replication within an operator-cell; the interaction estimate is therefore reported descriptively alongside the pre-specified paired cell contrasts.
\item \texttt{-Diagnostic} and \texttt{-Adaptive scheduling} are bundled ablations. Their outcomes cannot be assigned to one subcomponent.
\item \texttt{-Full-eval} is a matched shadow from the Full trajectory. It estimates the acceptance risk of stopping at the first compact pass, but is not an independent run.
\item The same final integrity criterion is applied symmetrically to every candidate; automatic anti-cheat and final verdicts are retained as separate evidence layers, together with available source-level resolution metadata.
\item Token fields are provider-reported execution volumes. Within-model ratios are comparable; cross-provider token totals and pricing proxies are not treated as auditable monetary cost.
\item The primary ablation cost uses the end-to-end task run that produces each operator's reported terminal outcome, following the predeclared task-level metric. Earlier runs terminated by infrastructure or provider interruptions are retained for stability diagnosis but excluded from the primary operator cost.
\end{enumerate}

\FloatBarrier
\section{S9. Summary of Supported Conclusions}

The extended evidence supports four calibrated conclusions:

\begin{enumerate}
\item A single state-continuous Debug-Agent task recovers 18/27 operators, compared with an average single-trial regeneration success of 21/81 and an operator-level three-trial Pass@3 of 11/27. It solves 11 operators that all regeneration trials miss, while Regenerate has four exclusive successes.
\item Debug-Agent reduces all tokens per Pass@1 by 92.8\% relative to the full three-trial regeneration budget. In the method \(\times\) model factorial, its all-token efficiency advantage over CANNBot precision-debug repeats within both Qwen and Kimi. CBD-Q has lower uncached-token volume per operator, while Debug-Agent has lower all-token volume per operator and per Pass@1 in both model strata.
\item The balanced \(2\times2\) factorial yields a +20.4 pp observed method main effect for Debug-Agent, a +13.0 pp model main effect for Kimi, and a -3.7 pp method-by-model interaction. Debug-Agent obtains 18/27 versus 12/27 Pass@1 outcomes under Qwen and 21/27 versus 16/27 under Kimi. The small observed interaction indicates a directionally stable method advantage across the two models, while the individual paired contrasts remain non-significant at \(p<0.05\).
\item The ablation evidence shows distinct outcome patterns when different capabilities are removed: removing retrieved knowledge is associated with the clearest decrease in recovery and token efficiency; removing diagnostic evidence corresponds to a weaker degradation; removing anti-cheat or full-coverage evaluation is associated with accepting unsupported successes; and removing the current adaptive-scheduling thresholds produces no measurable degradation on this cohort.
\end{enumerate}

\end{document}